\documentclass[12pt]{iopart}
\usepackage{graphicx}
\usepackage{dcolumn}
\usepackage{bm}

\begin{document}

\title{Effect of Sr substitution on superconductivity in
Hg$_{2}$(Ba$_{1-y}$Sr$_{y}$)$_{2}$YCu$_{2}$O$_{8 - \delta }$
(part~1): a neutron powder diffraction study}

\author{P Toulemonde\dag\footnote[3]{Present address: LPMCN, Universit\'{e} Lyon-I,
B\^{a}timent L\'{e}on Brillouin, 43 Boulevard du 11 Novembre 1918,
F-69622 Villeurbanne cedex, France.}, P Odier\dag, P Bordet\dag, S
Le Floch\dag and E~Suard\ddag}

\address{\dag\ Laboratoire de Cristallographie, CNRS, 25 avenue des martyrs,
BP166, F-38042 Grenoble cedex 09, France.}
\address{\ddag\ Institut Laue-Langevin, 6 rue Jules Horowitz, BP156,
F-38042 Grenoble cedex 09, France.}

\ead{pierre.toulemonde@lpmcn.univ-lyon1.fr}

\date{\today}

\begin{abstract}
The effect of Sr chemical pressure on superconductivity was
investigated in
Hg$_{2}$(Ba$_{1-y}$Sr$_{y}$)$_{2}$YCu$_{2}$O$_{8-\delta}$. The
samples were synthesized at high pressure -- high temperature from
y~=~0.0 to full substitution, y~=~1.0. These Sr-substituted
compounds are superconducting, without Ca doping on the Y site,
and show an increasing $T_{c}$ with Sr, reaching 42~K for y~=~1.0.
A detailed neutron powder diffraction study compares the
structural changes induced by this chemical Sr/Ba substitution and
the mechanical pressure effects in the Hg-2212 system. It shows
the strong decrease of the three Ba/Sr-O distances and
consequently the shrinkage of Cu-O1$_{in-plane}$ and
Cu-O2$_{apical}$ bonds. These structural changes, by affecting the
charge transfer which occurs between the charge reservoir and the
superconducting block, are responsible of the $T_{c}$ enhancement
with Sr content.
\end{abstract}

\submitto{\JPCM} 

\maketitle

\section{\label{sec:level1}Introduction}

The high pressure -- high temperature (HP-HT) process allows to
obtain nearly pure HgBa$_{2}$Ca$_{n-1}$Cu$_{n}$O$_{2 + 2n + \delta
}$ mercury based cuprate superconductors, even for
n~$>$~3~\cite{Capponi1,Loureiro0}. It is also efficient for the
synthesis of Hg$_{2}$Ba$_{2}$YCu$_{2}$O$_{8-\delta }$ compound
(Hg-2212)~\cite{Radaelli1,Radaelli2,Radaelli3}. In
Hg$_{2}$Ba$_{2}$YCu$_{2}$O$_{8-\delta }$ a strong oxygen
deficiency ($\delta $~$\sim $~0.4-0.5) inhibits superconductivity.
Ca$^{2 + }$ substitution on the Y$^{3 + }$ site in the series
Hg$_{2}$Ba$_{2}$(Y$_{1-x}$Ca$_{x})$Cu$_{2}$O$_{8-\delta } $
restores the superconducting state but the Ca solubility limit
(x~$\sim $~0.4) prevents to achieve the optimal doping regime.
Thallium (Tl$^{3 + })$ or lead (Pb$^{4 + })$ substitution for
Hg$^{2 + }$ allows to extend the Ca solubility domain and to
achieve a $T_{c~max}$ of
82-84~K~\cite{Tokiwa,Toulemonde1,Toulemonde2,Toulemonde3}.

The Hg-2212 compound shows a giant $T_{c}$ increase under
mechanical pressure, not only for underdoped but also for nearly
optimally doped samples: +~50~K under 20~GPa~\cite{Acha}. This
enhancement is the greatest ever measured in superconducting
compounds. In order to stabilize such possible $T_{c}$ increase at
ambient pressure~\cite{Marezio1,Marezio2,Marezio3}, ``chemical
pressure'' was attempted in cuprates by using Sr substitution for
Ba. Nevertheless in most cases -~except for the La$_{2}$CuO$_{4}$
system~\cite{Cava,Torrance,Radaelli4}, where the substitution is
not isovalent and changes the doping~- the replacement of Ba by Sr
produces a continuous decrease of
$T_{c}$~\cite{Wada,Ganguli,Subramian,Karpinski1,Sin}. Sr
substitution has been already tried in the double mercury layer
system, but the first attempts lead to an impure Hg-2212 sample
accompanied with Hg-1212~\cite{Loureiro,Chaillout1}.

The first aim of this study was to find the optimal experimental
conditions to obtain nearly pure
Hg$_{2}$(Ba$_{1-y}$Sr$_{y})_{2}$YCu$_{2}$O$_{8 - \delta }$
samples. Then the $T_{c}$ of these Hg-2212 samples was measured
versus Sr content, and surprisingly, an increase of $T_{c}$ was
observed with Sr. The $T_{c}$ increased from 0~K in the
unsubstituted sample (y~=~0) to 42~K for full substitution
(y~=~1.0). This enhancement was not only observed in the Y-based
Hg-2212 series but also confirmed in a second Hg-2212 series,
doped with 20~{\%} of Ca on the Y site, where $T_{c}$ increases
now from 21~K (y~=~0) to 58-60~K
(y~=~0.7-1.0)~\cite{Toulemonde4,Toulemonde5}. Finally, we
performed neutron powder diffraction (NPD) on both series to
better determine the origin of this $T_{c}$ enhancement.

This study is divided in two papers. In this first one, we discuss
the structure (from NPD) of two compositions of the Y-based series
(y~=~0.50 and y~=~1.0) which were carefully compared with that for
y~=~0.0~\cite{Radaelli1}. We also compare the effects induced by
Sr substitution with those of the mechanical pressure. For that, a
second set of structural informations obtained from refinements of
high pressure synchrotron X-ray diffraction patterns was used.
These data were acquired during a previous experiment carried out
on the high pressure ID-09 beamline at ESRF (European Synchrotron
Radiation Facility, Grenoble, France) for a Hg-2212 sample up to
30~GPa using a diamond anvil cell (for more details, see
ref.\cite{Loureiro,Bordet1}). We conclude to a positive chemical
pressure effect, showing the structural changes associated to the
$T_{c}$ increase with Sr.

In the second paper~\cite{ToulemondeBVS}, we have investigated the
charge repartition into the cell by "bond valence sum" analysis,
starting from the structure (determined by NPD). Different origins
of a charge transfer to the superconducting CuO$_{2}$ plane are
identified, which could explain the $T_{c}$ enhancement with Sr.

\section{\label{sec:level2}Experimental}

Two high pressure equipments were used to synthesize the samples.
The first syntheses were first carried out using a small belt
system (samples of $\sim $~500~mg) to select the best conditions
for the HP-HT synthesis. The pressure was kept at 3.5-4~GPa and
different time (15-120~min) and temperature (900-1050\r{ }C)
conditions were applied. Then, the HP-HT syntheses were carried
out in a large volume Conac HP system, with the optimized pressure
and temperature conditions, in order to produce larger samples
($\sim $~2.4~g) for NPD experiments.

Two different starting mixtures were used; the first one to select
the appropriate temperature (T), time (t) and pressure (P) of
synthesis to get the purest possible sample. We used the  belt
apparatus that consumes only a small amount of powder. The powders
were prepared by mixing simple oxides HgO (Aldrich, 99~{\%}),
BaO$_{2}$ (Merck, 95~{\%}), SrO$_{2}$ freshly prepared by
precipitation from a nitrate solution of strontium (Aldrich,
99~{\%}), Y$_{2}$O$_{3}$ (Prolabo, 99.9~{\%}), CuO (Aldrich,
99~{\%}) and metallic Cu (Ventron, 99~{\%}) in the appropriate
ratio to adjust the oxygen content. After selecting T, t and P,
larger batches were processed for NPD experiments in the Conac HP
cell. In this case, we used the second kind of mixture based on
precursors of ``(Ba$_{1-y}$Sr$_{y})_{2}$YCu$_{2}$O$_{z}$''
prepared according to the procedure described in
ref.~\cite{Toulemonde6}. More precisely, for y~=~0.50, two
different precursors ``Ba$_{2}$YCu$_{2}$O$_{z}$'' and
``Sr$_{2}$YCu$_{2}$O$_{z}$'' were prepared by heating the
corresponding mixture of oxides at 850~\r{ }C and 950\r{ }C
respectively for 24~h. To the precursor(s), HgO was added in the
stoichiometric ratio to get the compositions y~=~0.50 and y~=~1.0.
Both samples were prepared at 3.5~GPa and 1050\r{ }C for 60 and
30~min respectively.

$T_{c}$ was measured by a.c. susceptibility; it was performed at
119~Hz using a home-made apparatus working at low magnetic field
of 0.012~Oe, in the range 4.2~--~300~K.

The microstructure and composition of samples were investigated by
electron microscopy on a Philips CM-300 TEM and JEOL-840 SEM, both
equipped with a Kevex system for X-ray energy dispersive
spectroscopy (EDX) analysis. In the EDX analysis many regions and
grains of the same phase were examined and the compositions
averaged. Nevertheless, the EDX composition was calculated without
the use of standards (such as Y${_2}$BaCuO${_5}$ and HgS) which
would allow a more precise quantification of elements.

X-ray diffraction patterns (XRD) were collected using a powder
diffractometer (Siemens D-5000) working in transmission mode at
the wavelength $\lambda _{Cu,~K\alpha 1}$~=~1.54056~{\AA}
(K${_{\alpha 2}}$ Ni filtered). The neutron powder diffraction
experiments were performed at room temperature on the D2B
instrument of ILL (Institut Laue-Langevin, Grenoble, France). Each
powder sample was introduced into a vanadium can to be supported
into the neutron beam. The wavelength was tuned to 1.594~{\AA} and
the average collecting time was eight hours which was enough to
achieve satisfactory counting statistics. The detector has 64
cells spaced by 2.5\r{ } permitting to record the NPD pattern from
2$\theta $~=~6 to 160\r{ }with an angular resolution of 0.05\r{ }.

\section{\label{sec:level3}Results}

\subsection{Phase composition, unit cell and pressure effect}

\begin{figure}
\begin{center}
\includegraphics[width=\linewidth]{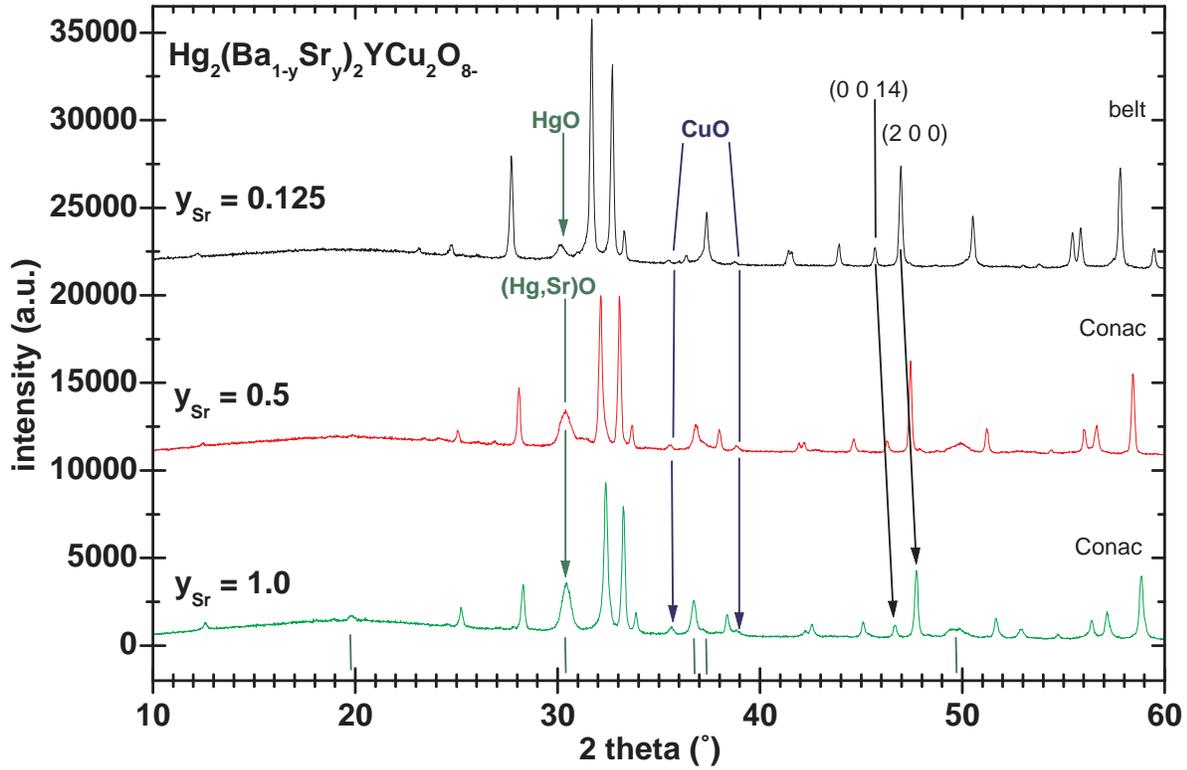}
\end{center}
\caption{\label{fig:eps1} XRD patterns for
Hg$_{2}$(Ba$_{0.875}$Sr$_{0.125})_{2}$YCu$_{2}$O$_{8 - \delta }$
(belt synthesis), Hg$_{2}$(Ba$_{0.5}$Sr$_{0.5})_{2}$YCu$_{2}$O$_{8
- \delta }$ and Hg$_{2}$Sr$_{2}$YCu$_{2}$O$_{8 - \delta }$ (Conac
synthesis) at $\lambda $~=~1.54056{\AA}.}
\end{figure}

The purity of samples, synthesized in the belt cell or in the
Conac HP system is almost identical, it does not depend on the
nature of the starting mixture. The main phase is Hg-2212, but
sometimes, for Sr rich compositions, Hg-1212 impurity phase is
observed due to some reaction with the gold capsule enhanced by Sr
content. To avoid the formation of Hg-1212 phase, the reaction
time should be reduced for Sr rich compositions (i.e. from 60~min
for y~=~0.5 to 30~min for y~=~1.0). Nevertheless, even with
optimized time conditions, impurities phases were observed
(figure~1): HgO, CuO and an unknown phase whose amount increases
with the nominal Sr content. This phase containing Hg, Sr and
oxygen has been reported to be Sr$_{0.76}$Hg$_{1.24}$O$_{2}$
according to X-ray diffraction (see tick marks in the bottom part
of figure~1) and EDX data~\cite{Toulemonde4,Toulemonde5}. We
assume this phase to have a unit cell close to that one of the
Ca$_{0.76}$Hg$_{1.24}$O$_{2}$ phase identified in Ca-rich
(Hg,Pb)-2212 compositions~\cite{Toulemonde3} and occurring also in
Hg-1223 HP-HT synthesis~\cite{LeFloch1,Bordet2}. The structural
model of this Sr$_{0.76}$Hg$_{1.24}$O$_{z}$ impurity (space group
I 4/mmm) will be used to calculate its contribution in the
Rietveld refinements (see paragraph 3.4.).

\begin{figure}
\begin{center}
\includegraphics[width=0.7\linewidth]{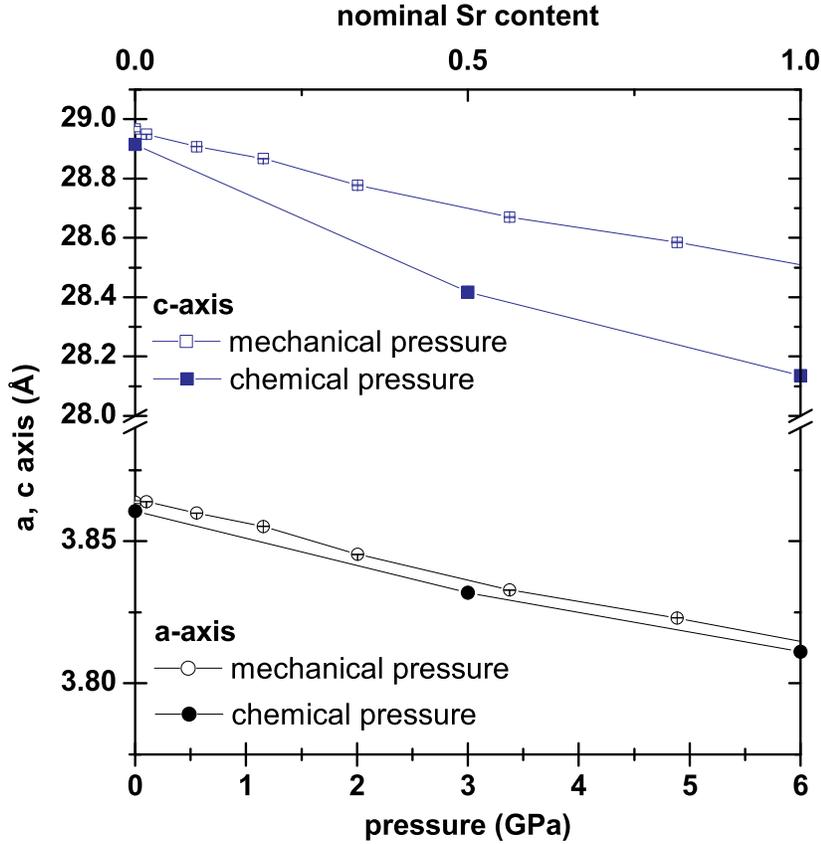}
\end{center}
\caption{\label{fig:eps2}Comparison of chemical and mechanical
pressure effects on the lattice parameters in Hg-2212; a and
c-axis are plotted versus nominal Sr content and versus pressure
respectively.}
\end{figure}

Table~1 shows the lattice parameters of our y~=~0.5 and 1.0
Hg-2212 samples, determined from NPD. They are compared to those
for y~=~0.0 (first column) from
Radaelli~et~al.~\cite{Radaelli1,Radaelli2,Radaelli3}. A
significant and anisotropic decrease of the unit cell is observed:
$\sim $~1.3~{\%} along the a-axis compared to $\sim $~2.7~{\%} for
c-axis. It shows the shrinkage of the lattice due to the
progressive substitution of Ba by Sr, with a drop of $\sim
$~5~{\%} of the unit cell volume. The comparison of a and c-axis
changes with Sr content (upper scale) and with mechanical pressure
(lower scale), previously obtained by synchrotron powder
diffraction~\cite{Loureiro,Bordet1}, is shown figure 2. We note
that both effects scale together (illustrated for a-axis)
suggesting that the substitution of 1 Ba for 1 Sr is equivalent to
the application of a pressure of $\sim $~6~GPa for a-axis and
$\sim $~10~GPa for c-axis. Nearly identical values were observed
by F.~Licci et al. in the Sr-substituted Y(Ba$_{1 -
x}$Sr$_{x})_{2}$Cu$_{3}$O$_{7 - \delta }$ system (i.e.
Cu-1212)~\cite{Licci}. In part~4 we will detail the atomic shifts
induced by the ``chemical pressure'' and compare them with those
observed by mechanical pressure.

\subsection{Electron microscopy investigations}

\begin{figure}
\begin{center}
\includegraphics[width=0.65\linewidth]{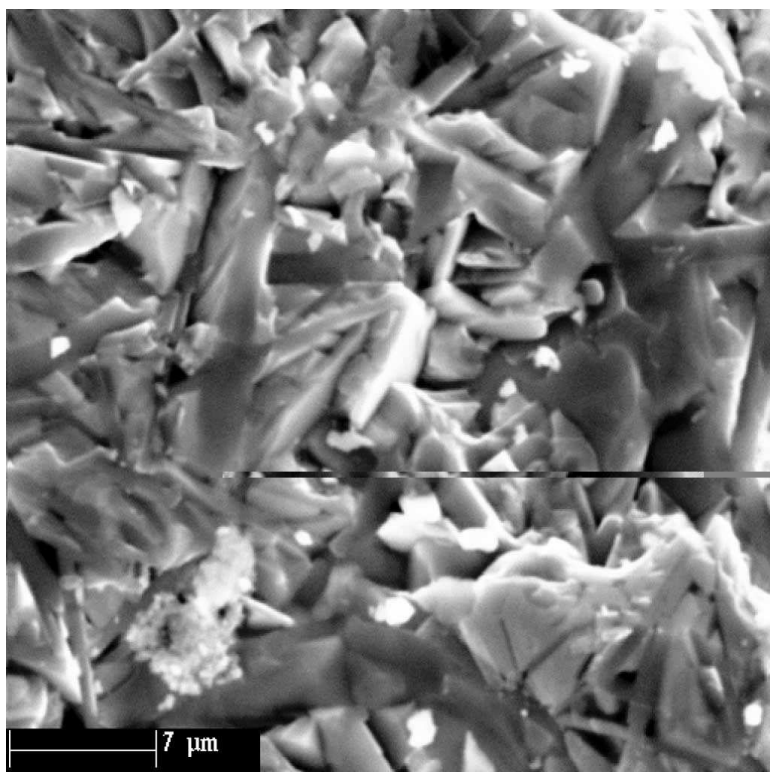}
\end{center}
\caption{\label{fig:eps3}Microstructure of y~=~1.0 Sr-substituted
sample (SEM image).}
\end{figure}

A sample of Hg$_{2}$Sr$_{2}$YCu$_{2}$O$_{8 - \delta }$ was
observed by scanning electron microscopy (SEM) and transmission
electron microscopy (TEM). Figure~3 shows its microstructure which
is typical of the series. It is composed of well defined platelets
of 5-10~$\mu $m long. The grains size is similar to that obtained
for (Hg,Pb)-2212 samples~\cite{Toulemonde3,Toulemonde7}. The
composition of the 2212 grains was found close to the nominal
stoichiometry by EDX. The impurity phase detected by XRD has also
been identified in SEM observations as small grains~$<$~1~$\mu $m.

\begin{figure}
\begin{center}
\includegraphics[width=0.5\linewidth]{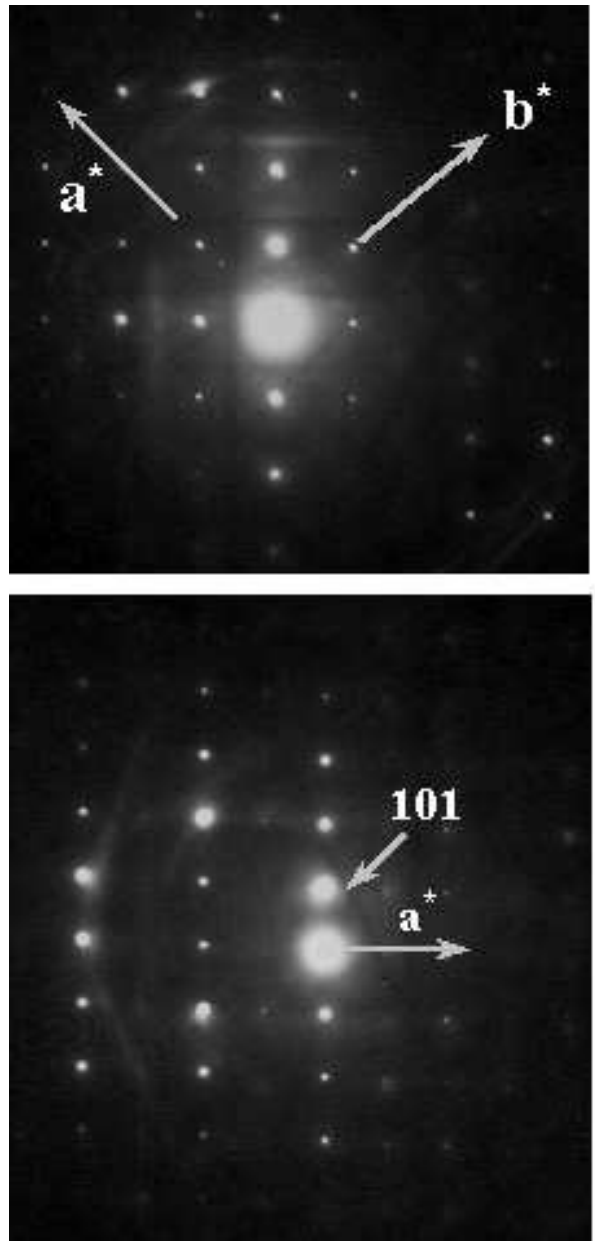}
\end{center}
\caption{\label{fig:eps4}Electron diffraction image in two
different orientations of typical crystallites of
Hg$_{2}$Sr$_{2}$YCu$_{2}$O$_{8 - \delta }$.}
\end{figure}

In figure~4 is shown the electron diffraction (ED) images obtained
using two different orientations of typical
Hg$_{2}$Sr$_{2}$YCu$_{2}$O$_{8 - \delta }$ crystallites. The
lattice parameters calculated from these ED images (a~=~3.81~{\AA}
and c~=~28.2 {\AA}) are consistent with those determined from XRD
or NPD patterns. The average composition of several individual
crystallites was determined by EDX using a nanoprobe of 20~nm. The
composition was
``Hg$_{1.42}$Sr$_{2.22}$Y$_{1.66}$Cu$_{2}$O$_{z}$'' which can be
interpreted as
``(Hg$_{0.71}$Y$_{0.30}$)$_{2}$(Sr$_{1.11}$)$_{2}$Y$_{1.06}$Cu$_{2}$O$_{z}$'',
suggesting the substitution of 30~{\%} of Y on the Hg site. This
substitution, not taken into account by earlier
reports~\cite{Radaelli1,Radaelli2,Radaelli3}, was clearly shown in
the next studies based on refinements of
Hg$_{2}$Ba$_{2}$YCu$_{2}$O$_{8 - \delta }$ NPD patterns
~\cite{Loureiro,Toulemonde4}. This information will be used in our
further refinements of the NPD patterns (paragraph 3.4.). The
10~{\%} excess Sr content might not be representative being within
the error of our EDX estimation.

\subsection{Superconducting properties}

As shown by a.c. susceptibility measurements both samples,
y~=~0.50 and y~=~1.0, are superconducting (see figure~5 in
ref.~~\cite{Toulemonde5}). This is surprising in view of the
absence of superconductivity generally found in
Hg$_{2}$Ba$_{2}$YCu$_{2}$O$_{8 - \delta
}$~\cite{Radaelli1,Radaelli2,Radaelli3}. Note that in (Hg,Tl)-2212
Tokiwa-Yamamoto et al.~\cite{Tokiwa} have observed
superconductivity at 21~K in the underdoped Tl-free compound
(superconductivity which disappears after post-annealing in
flowing~Ar). The superconducting volume fraction shown here ($
\approx $ 35-50~{\%}) is high enough to rule out the contribution
of another superconducting phase not detected by XRD. As a result,
the superconducting transitions at 32~K for y~=~0.5 and 42~K for
y~=~1 (diamagnetic onset) are unambiguously attributed to the Sr
substituted Hg-2212 phase. Hence $T_{c}$ increases from 0~K to
32~K for y~=~0.5 and 42 K for full substitution. This is the first
example in cuprates, except in
(La,Sr)$_{2}$CuO$_{4}$~\cite{Cava,Torrance,Radaelli4}, or in
artificially stressed epitaxial (La,Sr)$_{2}$CuO$_{4}$
films~\cite{Locquet}, where the Sr substitution rises strongly
$T_{c}$. The broadening of the transition is probably due to
sample inhomogeneities.

As mentioned in introduction, we have recently confirmed this
$T_{c}$ enhancement with Sr chemical pressure in the Hg-2212
system in a second series, doped with 20~{\%} of Ca on the Y site,
i.e. with a higher doping level. In this second series, $T_{c}$
increases from 21~K (y~=~0) to 58-60~K
(y~=~0.7-1.0)~\cite{Toulemonde4,Toulemonde5}. Then, the phenomenon
of $T_{c}$ increase seems to be intrinsically due to Sr.

\subsection{Structure: Rietveld refinements}

The structural parameters of y~=~0.5 and 1.0 samples were refined
by the Rietveld method using the software ``fullprof''. Only the
NPD data in the  2$\theta $ range 10-160\r{} were taken into
account. The profile shape was calculated based on a pseudo-Voigt
function. The background was approximated by a sixth order
polynomial. The absorption was evaluated and corrected using a
cylinder approximation of the sample shape.

\begin{figure}
\begin{center}
\includegraphics[width=0.3\linewidth]{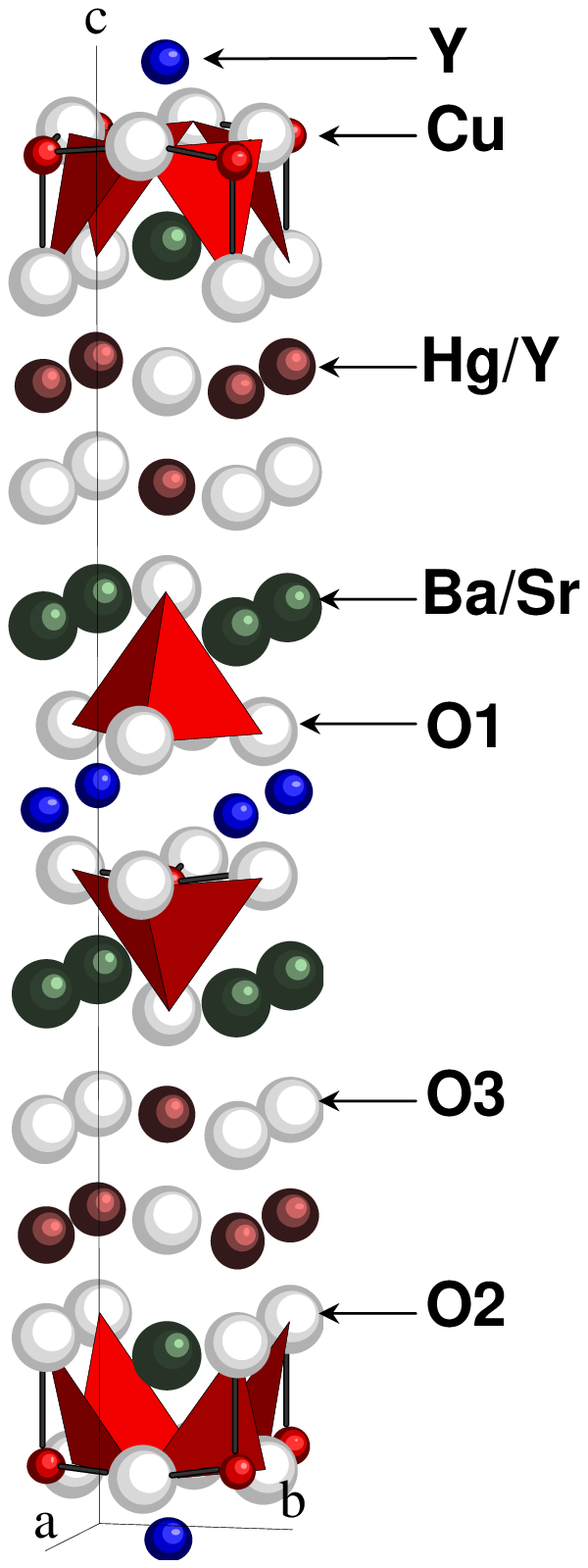}
\end{center}
\caption{\label{fig:eps3}Schematic structure of the Hg-2212
lattice (space group I/4 m m m) used for NPD Rietveld
refinements.}
\end{figure}

The Hg-2212 structure (illustrated in figure~5) was refined based
on the tetragonal model (space group I 4/mmm) previously
established for Ba-based compounds by
Radaelli~et~al.~\cite{Radaelli1,Radaelli2,Radaelli3}: Hg/Y at 4e
(0, 0, z~$\sim $~0.21), Ba at 4e (1/2, 1/2, z~$\sim $~0.13), Y at
2b (1/2, 1/2, 0), Cu at 4e (0, 0, z~$\sim $~0.055), O1 at 8g (1/2,
0, z~$\sim $~0.05), O2 at 4e (0, 0, z~$\sim $~0.14) and O3 at 4e
(1/2, 1/2, z~$\sim $~0.22). The only difference in our
calculations consists in taking into account the Sr substitution
for Ba but also a possible substitution of Y on Hg site, as shown
by S.M. Loureiro in Hg$_{2}$Ba$_{2}$YCu$_{2}$O$_{8 - \delta
}$~\cite{Loureiro} and confirmed in our previous
work~\cite{Toulemonde4}. All the Debye Waller factors were chosen
isotropic. We tested anisotropic factors for Hg/Y and O3 sites but
they did not improve the refinements.

\begin{figure}
\begin{center}
\includegraphics[width=0.9\linewidth]{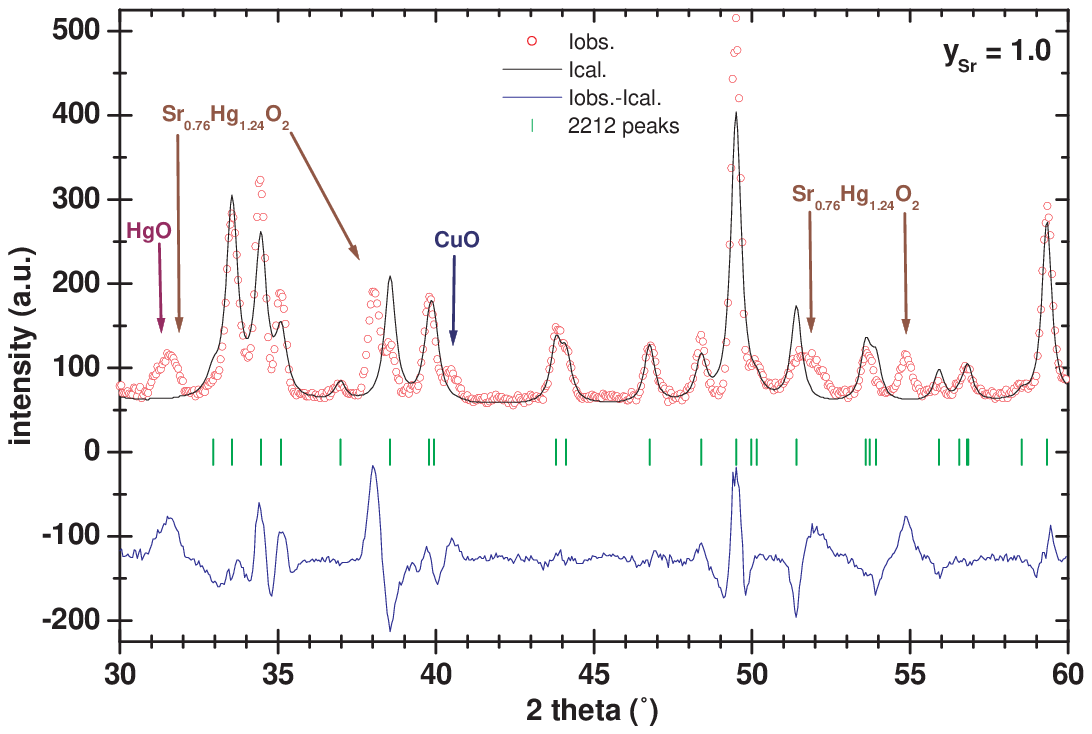}
\end{center}
\caption{\label{fig:eps6}Rietveld refinement profile of neutron
powder diffraction data ($\lambda $~=~1.594~{\AA}) for
Hg$_{2}$Sr$_{2}$YCu$_{2}$O$_{8 - \delta }$ at room temperature
taking into account only the Hg-2212 phase. A difference curve is
plotted at the bottom (observed minus calculated). It shows the
contribution of HgO, CuO and Sr$_{0.76}$Hg$_{1.24}$O$_{2}$
secondary phases.}
\end{figure}

The secondary phases were taken into account but only their scale
factors, lattice parameters and profile shape parameters were
refined. Figure~6 shows their contribution to the NPD diffracted
intensity in the 2$\theta $ range 30-60\r{}. The merit factors of
the refinements were significantly improved by taking into account
the HgO and CuO impurities, but also the Sr-Hg-O phase observed by
XRD and EDX. The structural model used was that of
Ca$_{0.76}$Hg$_{1.24}$O$_{2}$ (space group I~4/mmm), established
previously~\cite{Bordet2}. In the present case, Ca was assumed to
be simply replaced by Sr. We tried also to partially replace Sr by
another element (Cu for example), but this partial substitution
had no measurable impact on the refined structural parameters
(atomic positions or Debye-Waller factors) of the main Hg-2212
phase.

\begin{figure}
\begin{center}
\includegraphics[width=0.7\linewidth]{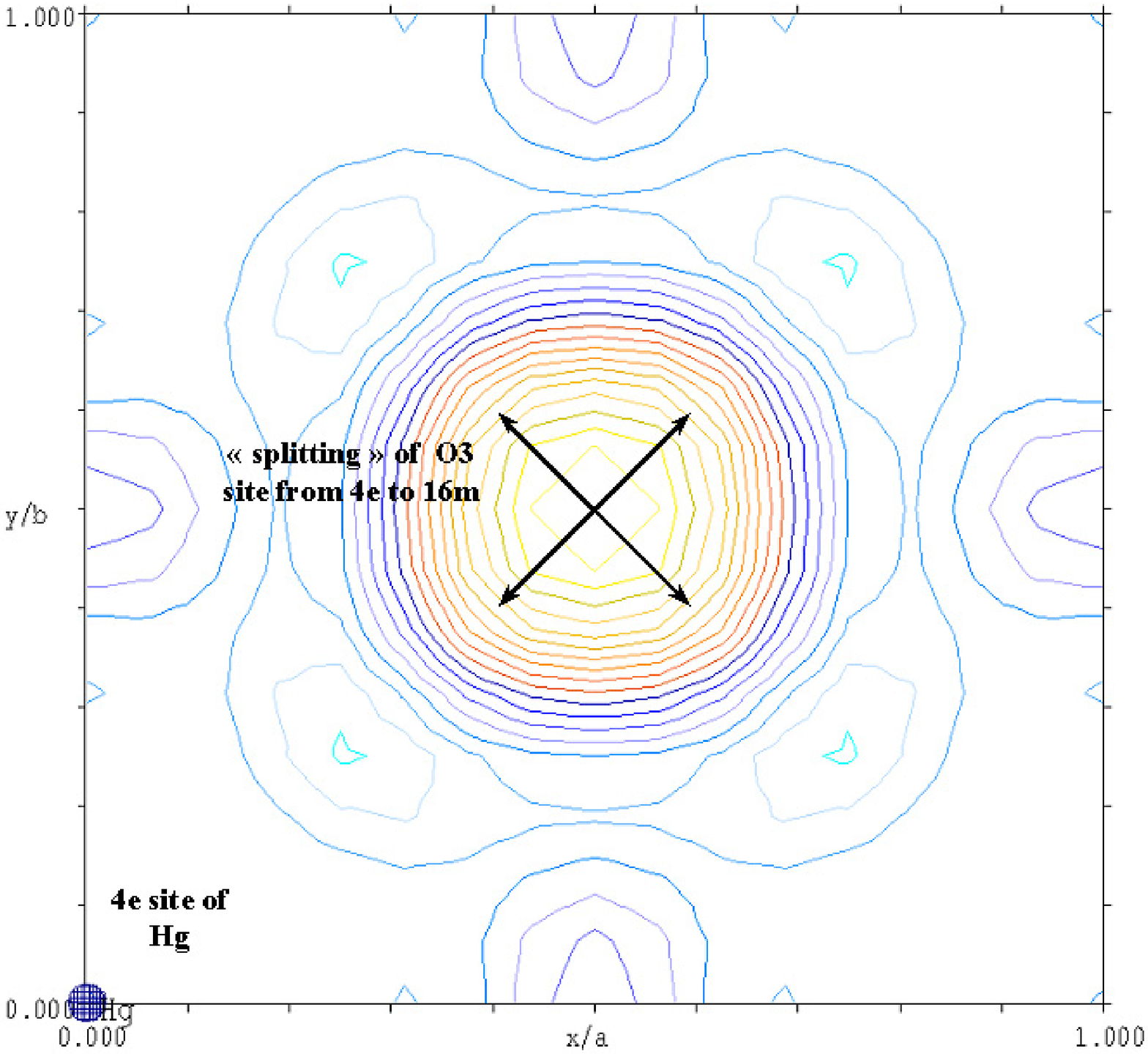}
\end{center}
\caption{\label{fig:eps7}Difference Fourier map
(F$_{obs}$~--~F$_{calc})$ of the nuclear density generated for the
(x,~y,~z~=~0.2105) section (i.e.~ (Hg,Y)O3 plane) of
Hg$_{2}$Sr$_{2}$YCu$_{2}$O$_{8 - \delta }$. The contribution of
the O3 oxygen was omitted from the calculations of F$_{calc}$ to
highlight the splitting.}
\end{figure}

In the starting model, the occupancy factors were supposed full.
All the thermal Debye-Waller factors converged to reasonable
values, except that for O3, in site 4e, which remains close to
5~{\AA}$^{2}$, suggesting a non-optimized position of O3 oxygen.
This is confirmed by the Fourier Difference map calculated from a
model without any O3 oxygen (figure~7). It indicates that O3 is
displaced from its 4e site towards a neighboring 16m site. The
absence of any additional O4 oxygen on the cell edge (in (0,~y,~z)
site) of the (Hg,Y)O3 plane was also checked. The position and the
occupancy factor of O3 were refined in its 16m (x, x, z)
configuration. The final values of Debye-Waller factors of O3
oxygen, did not decreased below 2~{\AA}$^{2}$, it denotes the
defective character of the (Hg,Y)O3 layer.

\begin{figure}
\begin{center}
\includegraphics[width=0.9\linewidth]{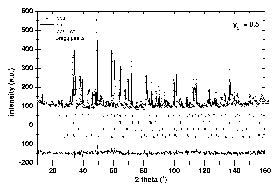}
\end{center}
\caption{\label{fig:eps8}Rietveld refinement profile of NPD data
($\lambda $~=~1.594~{\AA}) for
Hg$_{2}$(Ba$_{0.5}$Sr$_{0.5})_{2}$YCu$_{2}$O$_{8 - \delta }$ at
room temperature. A difference curve is plotted at the bottom
(observed minus calculated). Tick marks correspond to Bragg peaks
of Hg-2212 phase, Sr$_{0.76}$Hg$_{1.24}$O$_{2}$, CuO and HgO.}
\end{figure}

The presence of Y in substitution to Hg, evidenced by EDX, has
been taken into account and the occupancy factor Y/Hg was refined.
The difference between scattering lengths of Y (0.775$\times
$10$^{ - 12}$~cm) and Hg (1.266$\times $10$^{ - 12}$~cm) was
sufficient to allow this calculation.

Among other substitution, Sr on Y site would have a strong impact
on doping in CuO${_2}$ planes. Unfortunately, NPD would not be
useful because of the too close values of their Fermi lengths
(b~=~0.702$\times $10$^{ - 12}$~cm for Sr) to make relevant
refinements of the occupancy of Y by Sr. However, the large ionic
radius difference between Sr${^{2+}}$ and Y${^{3+}}$ (1.26~{\AA}
compared to 1.019~{\AA} in coordination number 8) make the
substitution rather improbable. Moreover, in
Hg(Ba,Sr)$_{2}$Ca$_{2}$Cu$_{3}$O$_{8 + \delta }$ where Ca has the
same environment as Y, and where the contrast in b values between
Sr and Ca exists, it was shown that Sr do not substitute the Ca
site~\cite{Chmaissem1}. Finally, this substitution is not seen in
other cuprates (Cu-1212~\cite{Licci} or
Cu-2212~\cite{Karpinski2}). We conclude that substitution of Y by
Sr is very improbable

It was also possible to refine the Sr content in substitution to
Ba. A distinct site for Y and Hg in the (Hg,Y)O3 plane was also
considered. Both cations could be in 4e, 16m (x, x, z) or 16n (0,
y, z) site, independently or not. Each case was tested but no
improvement of the refinement was found. Then, we chose the
simplest case as the most representative one, this is when Y and
Hg are on a common 4e site. On figure 8 are drawn the observed,
calculated and difference curves of the
Hg$_{2}$(Ba$_{0.5}$Sr$_{0.5})_{2}$YCu$_{2}$O$_{8 - \delta }$
pattern after refinement; it shows an excellent agreement between
the experimental and the calculated profile.

\section{\label{sec:level4}Discussion: effects of chemical pressure on the structure}

The discussion is organized in two parts. In the first one, we
discuss the detailed effects of the Sr substitution on the atomic
positions. Doing that we extract the chemical pressure effect that
will be compared to that of the mechanical pressure.

Tables~1 and~2 show the refined parameters for y~=~0.50 and 1.0.
The structure of the y~=~0.0 compound (unsubstituted), determined
by Radaelli et al.~\cite{Radaelli1,Radaelli2,Radaelli3}, has been
added for comparison.

\begin{table}
\caption{\label{tab:table1} $T_{c}$ and refined structural
parameters from neutron powder diffraction data for
Hg$_{2}$Ba$_{2}$YCu$_{2}$O$_{8 - \delta }$,
Hg$_{2}$(Ba$_{0.5}$Sr$_{0.5})_{2}$YCu$_{2}$O$_{8 - \delta }$ and
Hg$_{2}$Sr$_{2}$YCu$_{2}$O$_{8 - \delta }$.}
\begin{indented}
\lineup
\item[]\begin{tabular}{@{}lcccc}
\br \multicolumn{2}{c}{Sr content} &y~=~0.0$^{\rm a}$ &y~=~0.5
&y~=~1.0\\ \mr
 a = b & ({\AA})& 3.8606(1)& 3.83195(5)& 3.8112(1) \\
 c & ({\AA})& 28.915(1)& 28.4176(7)& 28.136(1) \\ \mr
  Y& B({\AA}$^{2})$& 0.3(1)& 0.74(8)& 0.6(1) \\ \mr
 Ba/Sr& z& 0.1252(2)& 0.1266(1)& 0.1264(2) \\
 & B({\AA}$^{2})$& 0.5(1)& 1.28(9)& 1.56(8) \\
 & n(Ba)& 1& 0.30(5)& - \\
 & n(Sr)& -& 0.70(5)& 1 \\ \mr
 Hg/Y& z& 0.2126(1)& 0.2125(1)& 0.2124(1) \\
 & B({\AA}$^{2})$& 1.46(6)& 1.98(6)& 1.86(7) \\
 & n(Hg)& 1& 0.83(2)& 0.76(2) \\
 & n(Y)& -& 0.17(2)& 0.24(2) \\ \mr
 Cu& z& 0.0560(1)& 0.0589(1)& 0.0592(1) \\
 & B({\AA}$^{2})$& 0.30(6)& 0.67(5)& 0.32(6) \\ \mr
 O1& z& 0.0495(1)& 0.05059(8)& 0.0517(1) \\
 & B({\AA}$^{2})$& 0.62(6)& 0.77(5)& 0.71(6) \\ \mr
 O2& z& 0.1414(3)& 0.1403(2)& 0.1400(2) \\
 & B({\AA}$^{2})$& 1.5(1)& 1.40(7)& 1.54(9) \\ \mr
 O3& x = y& 0.574(2)& 0.589(2)& 0.576(2) \\
 & z& 0.2163(4)& 0.2150(4)& 0.2124(4) \\
 & B({\AA}$^{2})$& 0.7(4)& 2.5(3)& 1.5(3) \\
 & n& 0.78(3)& 0.94(2)& 0.90(2) \\ \mr
 R$_{wp}$& & 4.99{\%}& 4.66{\%}& 6.04{\%} \\
 Bragg R-factor& & not available& 7.20{\%}& 8.31{\%} \\
 $\chi ^{2}$& & 1.070& 2.54& 3.35 \\ \mr
 $T_{c}$ onset& & 0~K& 32~K& 42~K \\
 \br
\end{tabular}
\item[] $^{\rm a}$ The version of the structure presented here for
y~=~0 is the original model determined by P.G.~Radaelli et
al.~\cite{Radaelli1,Radaelli2,Radaelli3}.
\end{indented}
\end{table}

Both Sr-doped samples (y~=~0.5, 1.0), have a similar refined O3
occupancy (0.94(2) and 0.90(2) respectively), i.e. close to one.
This nearly full O3 occupancy is a consequence of the presence of
Y$^{3 + }$ substituting Hg$^{2 + }$ (17~{\%} and 24~{\%} for
y~=~0.5 and 1.0 respectively, table~1), filling oxygen vacancies
in the (Hg,Y)O3 plane. The presence of Y on the Hg site agrees
with SEM data with a refined value, 0.17(2) for y~=~0.50 and
0.24(2) for y~=~1.0 sample respectively, that is close to that
deduced from EDX ($ \approx $~0.30 for y~=~1.0 sample). Let us
recall that S.M.~Loureiro~\cite{Loureiro} refined also the
original NPD data for y~=~0.0 of Radaelli et
al.~\cite{Radaelli1,Radaelli2,Radaelli3} using a rather complex
model. His conclusion tends to prove that Y substitutes partially
(13~{\%}) the Hg site and provides a high occupancy of O3 (0.88(1)
instead of 0.78(3), i.e. corresponding to the oxygen composition
O7.76). We have confirmed this suggestion by synthesizing
(Hg$_{0.85}$Y$_{0.15}$)$_{2}$Ba$_{2}$YCu$_{2}$O$_{8}$ (nominal
composition)~\cite{Toulemonde4}. The refinement of Y occupancy on
Hg site from NPD data gives a value of 21~{\%} that is consistent
with the synthesis. Then, the O3 occupancy for y~=~0.5 and 1.0
compositions is finally close to that of pure Ba-based sample:
0.88(1), as recalculated by S.M.~Loureiro. The refined occupancy
factor of Sr for y~=~0.50 sample, 0.70(5), is probably
overestimated.

From the refined compositions (O3 and Y/Hg occupancies), one can
calculate the formal copper valence; it gives 2.13 (with 13~{\%}
of Y on the Hg site and n(O3)~=~0.88), 2.21 and 2.06 respectively
for y~=~0.0, 0.5 and 1.0. No coherent correlation between this Cu
valence (correlated to small oxygen content variations around
O$_{7.76}$-O$_{7.88}$ and to the partial Y content on the Hg site)
and the continuous increase of T$_{c}$ can be done. Moreover, as
shown by Alonso et al.~\cite{Alonso1}, a simple formal valence
analysis is not appropriate in the case of Hg-2212. The doping
level of the CuO${_2}$ planes should be lower than expected from
ionic considerations.

\begin{figure}
\begin{center}
\includegraphics[width=0.5\linewidth]{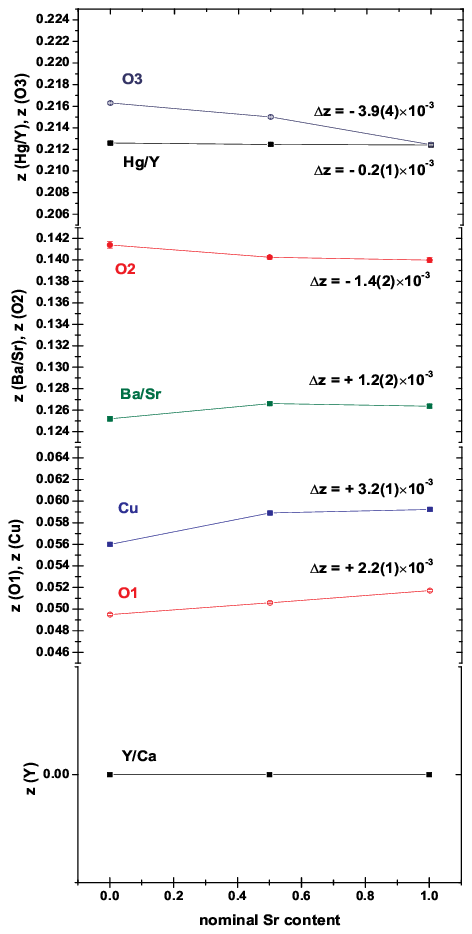}
\end{center}
\caption{\label{fig:eps9}z-position of atomic sites in
Hg$_{2}$Ba$_{2}$YCu$_{2}$O$_{8 - \delta
}$~\cite{Radaelli1,Radaelli2,Radaelli3},
Hg$_{2}$(Ba$_{0.5}$Sr$_{0.5})_{2}$YCu$_{2}$O$_{8 - \delta }$ and
Hg$_{2}$Sr$_{2}$YCu$_{2}$O$_{8 - \delta }$ as a function of the
nominal Sr content.}
\end{figure}

Figure~9 shows the effect of the Sr (nominal) content on
z-coordinate of the different atomic sites. The graphs are stacked
vertically as in the real structure with the same scale for direct
comparison. The total variation ``$\Delta $z'' is indicated on the
figure. Although the refined positions are close to those of the
original 2212 model without strontium, some changes are clearly
observed. The most striking effect concerns Cu and O1 atoms of the
superconducting block ``SB'' (composed of the two CuO$_{2}$
superconducting planes separated by the Y~plane) whose
z-coordinates increase respectively by 3.2$\times $10$^{ - 3}$ and
2.2$\times $10$^{ - 3}$ when Sr content increases from y~=~0 to
y~=~1. This change has no effect on the buckling angle of the
superconducting planes which remains close to $\sim $~13-14 deg.
(table~2) while the cell shrinks significantly. The other atoms
(Ba/Sr, O2, Hg/Y) do not move more than 1.4$\times $10$^{ - 3}$.
The O3 oxygen of the (Hg,Y)O3 plane is displaced towards the Hg/Y
plane by 3.9$\times $10$^{ - 3}$, making the (Hg/Y)O3 plane flat.
At the same time, the O3 atom shifts in the direction of Ba/Sr
site, and consequently, the (Ba/Sr)-O3 bond length decreases.

\begin{table}
\caption{\label{tab:table2}Selected interatomic distances
(in~{\AA}) and bond angles (in degrees) for
Hg$_{2}$Ba$_{2}$YCu$_{2}$O$_{8 - \delta
}$~\cite{Radaelli1,Radaelli2,Radaelli3},
Hg$_{2}$(Ba$_{0.5}$Sr$_{0.5})_{2}$YCu$_{2}$O$_{8 - \delta }$ and
Hg$_{2}$Sr$_{2}$YCu$_{2}$O$_{8 - \delta }$ as obtained from
Rietveld refinement of NPD.}
\begin{indented}
\item[]\begin{tabular}{@{}llll}
\br
Sr~content&\mbox{y~=~0.0}&\mbox{y~=~0.5}&\mbox{y~=~1.0}\\
\mr
 Y-O1& 2.403(2) & 2.395(1) & 2.398(2) \\ \mr
 Ba/Sr-O2& 2.770(2 )& 2.737(9) & 2.722(1) \\
 Ba/Sr-O1& 2.919(5) & 2.888(3) & 2.836(4) \\
 Ba/Sr-O3& 2.66(1) & 2.56(1) & 2.46(1) \\ \mr
 Cu-O1& 1.9396(5) & 1.9305(5) & 1.9173(5) \\
 Cu-O2& 2.469(9) & 2.312(6) & 2.272(8) \\ \mr
 Hg/Y-O2& 2.057(7) &2.052(6) &2.038(7) \\
 Hg/Y-O3$^{\rm b}$ & 2.10(1) & 2.116(1) & 2.15(1) \\
 Hg/Y-O3$^{\rm c}$ & 2.328(9) & 2.231(6) & 2.284(6) \\
 Hg/Y-O3$^{\rm c}$ & 2.762(2) & 2.753(7) & 2.726(6) \\
 Hg/Y-O3$^{\rm c}$ & 3.136(9) & 3.190(6) & 3.106(6) \\ \mr
 Cu-O1-Cu& 168.8(3) & 165.95(2) & 167.33(2) \\ \br
\end{tabular}
\item[] $^{\rm b}$ vertical.
\item[] $^{\rm c}$ in-plane.
\end{indented}
\end{table}

\begin{figure}
\begin{center}
\includegraphics[width=0.50\linewidth]{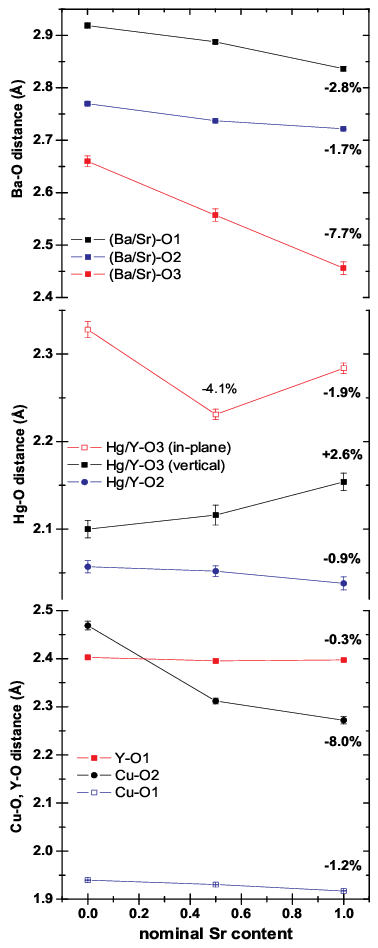}
\end{center}
 \caption{\label{fig:eps10}Interatomic distances
(in~{\AA}) for Hg$_{2}$Ba$_{2}$YCu$_{2}$O$_{8 - \delta
}$~\cite{Radaelli1,Radaelli2,Radaelli3},
Hg$_{2}$(Ba$_{0.5}$Sr$_{0.5})_{2}$YCu$_{2}$O$_{8 - \delta }$ and
Hg$_{2}$Sr$_{2}$YCu$_{2}$O$_{8 - \delta }$ samples as a function
of the nominal Sr content. The label indicates the total shrinkage
in~{\%}.}
\end{figure}

The modifications of the ``(Hg/Y)O$_{6}$'', ``(Ba/Sr)O$_{9}$'',
``YO$_{8}$'' and ``CuO$_{5}$'' coordination polyhedra are
illustrated in figure~10 with the same scale. The principal effect
is observed for the three neighboring oxygen O1, O2 and O3 of the
Ba/Sr site which become closer (along c axis) to the Ba/Sr site,
as a consequence of its smaller mean ionic radius. However the
shrinkage is not isotropic. The most significant change concerns
O3 atoms that become 7.7~{\%} closer to Ba/Sr while Ba/Sr-O1
shrinks by 2.8~{\%} and Ba/Sr-O2 only by 1.7~{\%} (Ba/Sr and O2
are located approximately at the same z). It is then obvious that
the distance connecting the SB together decreases significantly.
Here, the chemical pressure does not mimic the mechanical pressure
of 1~GPa which compresses Ba-O1, lets Ba-O2 unchanged and extends
Ba-O3 bonds in Hg-2212 (NPD data)~\cite{Loureiro,Bordet1}. In the
mono-Hg-layer family, the compressibility of bonds is extremely
sensitive to the doping level, i.e. the O3 oxygen concentration in
the mercury plane. The behavior of Ba-O bonds in our
Sr-substituted Hg-2212 is close to that observed in overdoped
Hg-1201 under mechanical pressure~\cite{Balagurov1,Antipov1}.

The influences of Sr substitution, i.e. ``chemical pressure'', are
not restricted to the Ba/Sr site, they extend beyond, up to the Cu
site. The distance Cu-O2 (apical) is the most significantly
renormalized bond by the Sr substitution. It amounts to a total
shrinkage of 8~{\%} (for y~=~1.0). The Cu-O2 distance becomes very
small, i.e. 2.27~{\AA}, compared to the value found in
Hg-12(n-1)n, i.e.
2.75~{\AA}-2.80~{\AA}~\cite{Antipov1,Radaelli5,Chmaissem2}. This
should affect the charge distribution and particularly the charge
transfer from the charge reservoir ``CR'' (composed by the double
(Hg/Y)O3 layer, linked to its two neighboring Ba/SrO2 planes) to
the SB. In agreement, such decrease is also observed in
Hg-12(n-1)n when holes are injected in the superconducting plane
by oxygen doping~\cite{Wagner1,Aksenov1}, but also in (La$_{1 -
x}$Ca$_{x})$(Ba$_{1.75 - x}$La$_{0.25 + x})$Cu$_{3}$O$_{y}$ where
moreover a maximum of the buckling angle occurs exactly at
T$_{c~max}$~\cite{Chmaissem3}. However, for n~=~1, it has been
demonstrated (see ref. in~\cite{Antipov1}) by fluorination that
Cu-O2 distance reduction does not cause an enhancement of T$_{c}$,
it is more likely the compression of the in-plane Cu-O1 distance
that is responsible of the T$_{c}$ enhancement. This goes in the
same way than that observed for strained films. In plane
compressive strain in epitaxial La$_{1.9}$Sr$_{0.1}$CuO$_{4}$ thin
films induces a doubling of T$_{c}$ from 25~K to 49~K while a-axis
(then Cu-O1 distance) decreases from 3.784~{\AA} (bulk) to
3.76~{\AA}~\cite{Locquet}, giving dT$_{c}$/da $\approx$
1000~K.{\AA}${^{-1}}$. In our Sr-substituted Hg-2212 samples,
Cu-O1 shrinks of only 1.2~{\%}, that is small, but follows the
trend. Interestingly, the corresponding T$_{c}$ increase rate
reaches 42/(3.8606~-~3.8112)~$\approx$~850~K.{\AA}${^{-1}}$.

In contrast, the Hg/Y-O2 distance shows smaller relative
variation, less than 1~{\%}. This is consistent with the smaller
compressibility observed for the Hg-O2 bond in
Hg-1201~\cite{Hunter}. Along the z-axis, O3 moves away from Hg/Y,
increasing the distance Hg/Y-O3 (between two adjacent Hg/YO3
planes) by 2.6~{\%}, which is similar to the decrease of the
(Ba/Sr)-O1 distance. This is the second highest effect, after the
shrinkage of (Ba/Sr)-O3. The variation of the in-plane Hg/Y-O3
distance, highly sensitive to the small changes of the Y content
on the Hg site, is not continuous and therefore can not be
correlated to the T$_{c}$ enhancement.

\begin{figure}
\begin{center}
\includegraphics[width=0.65\linewidth]{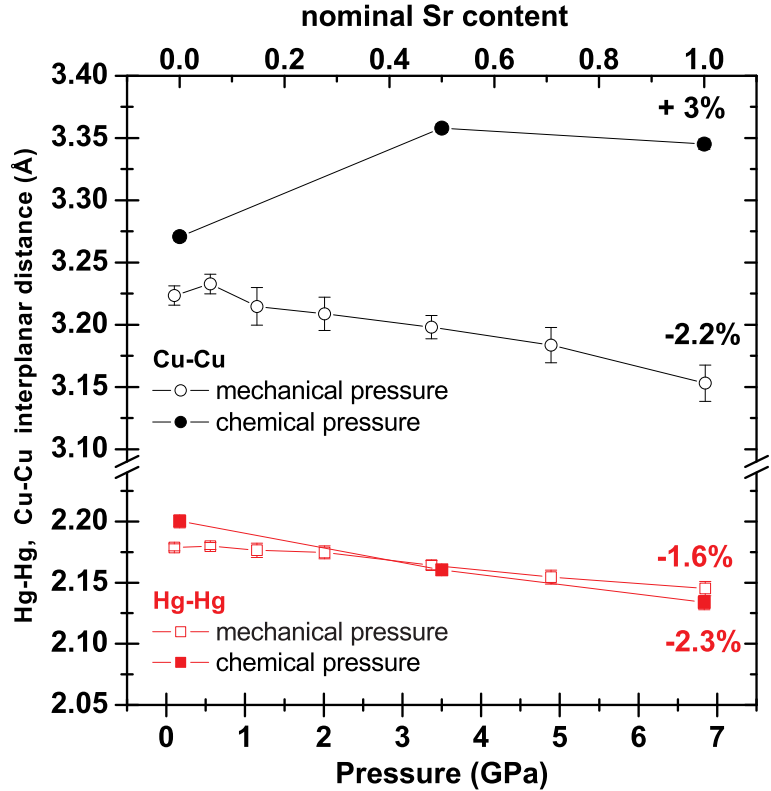}
\end{center}
\caption{\label{fig:eps11}Comparison of chemical and mechanical
pressure effects on the internal SB and CR thickness in Hg-2212:
Cu-Cu and Hg-Hg distances versus nominal Sr content and versus
pressure respectively.}
\end{figure}

\begin{figure}
\begin{center}
\includegraphics[width=0.65\linewidth]{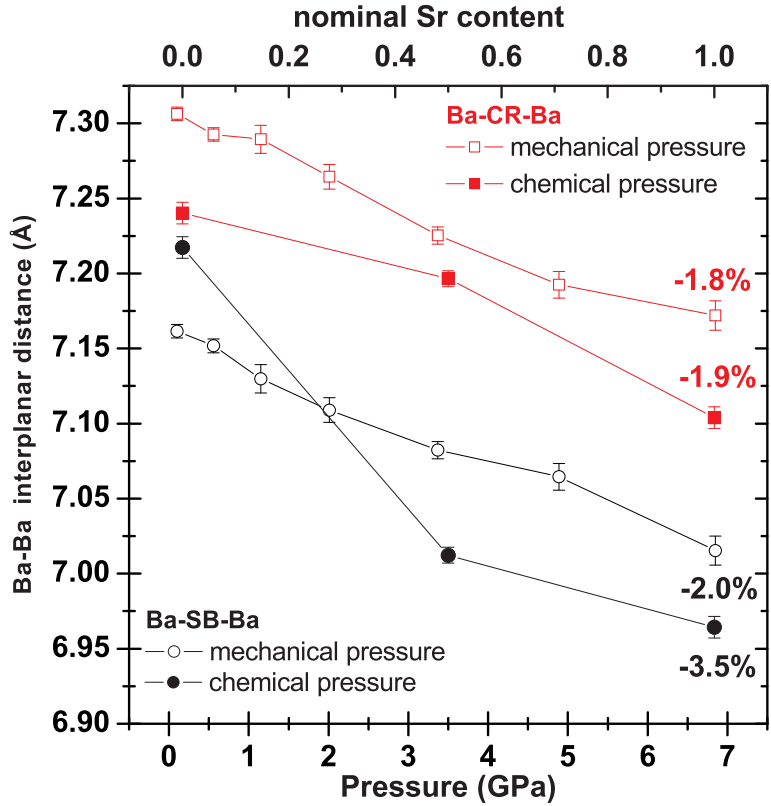}
\end{center}
\caption{\label{fig:eps12}Comparison of chemical and mechanical
pressure effects on the external SB and CR thickness in Hg-2212:
Ba-SB-Ba and Ba-CR-Ba distances versus nominal Sr content and
versus pressure respectively.}
\end{figure}

It is now interesting to compare the effects of chemical pressure
and mechanical pressure~\cite{Loureiro,Bordet1}. The chemical
pressure for y~=~1.0 is equivalent to a mechanical pressure of
6~GPa approximatively, as said in part~3.1. We will restrict the
analysis to pressures lower than 7~GPa. Figure~11, and~12, compare
the variations of the CR and SB thicknesses and, Hg-Hg and Cu-Cu
distances, for chemical pressure (NPD data) and mechanical
pressure (synchrotron data). They are similar, except one feature:
the Cu-Cu distance which quantifies the SB thickness. It increases
by 3~{\%} with Sr substitution, but decreases by~2~{\%} by
applying mechanical pressure. The same effect is observed in the
monolayer (Hg,Re)-1223 family: the complete substitution of Sr
increases the SB thickness by 3~{\%}~\cite{Armstrong,Marezio4}.
The same trend is also observed in the Cu$_{2}$(Ba$_{1 -
x}$Sr$_{x})_{2}$YCu$_{2}$O$_{8}$~\cite{Karpinski2} and Cu(Ba$_{1 -
x}$Sr$_{x})_{2}$YCu$_{2}$O$_{7 - \delta }$~\cite{Licci} compounds.
This increase is then intrinsic to the Sr presence, it cannot be
related to $T_{c}$ which increases with Sr substitution in Hg-2212
while it decreases in the latter ones. At the same time, the
interplanar Ba-SB-Ba thickness decreases drastically by 3-4~{\%}.
Then, the BaO2 and CuO1 planes move towards each others, which
leads to the great decrease of the apical Cu-O2 distance to
2.27~{\AA}. The variation of the Hg-Hg distance (-2.3~{\%})
follows the variation of the Ba-CR-Ba thickness which decreases by
about 2~{\%}.

\section{\label{sec:level5}Conclusion}

The structural effect of chemical pressure in
Hg$_{2}$(Ba$_{1-y}$Sr$_{y})_{2}$YCu$_{2}$O$_{8-\delta}$ was
studied by neutron powder diffraction and correlated to the
superconducting properties. The Sr substitution in Hg-2212 has a
positive effect on $T_{c}$ contrarily to the majority of cuprates.
It increases gradually $T_{c}$ from 0~K (y~=~0.0) to
42~K~(y~=~1.0). Moreover the Sr substituted compounds are already
superconducting without Ca-doping on the Y site. No coherent
correlation can be made between the continuous $T_{c}$ increase
and the variation of the formal valence of Cu in the Sr
substituted series.

From a structural point of view, the chemical pressure introduced
by Sr is similar to the mechanical pressure, except for the SB
thickness which increases with Sr substitution while it decreases
under external pressure. However, this is not correlated with the
variation of $T_{c}$, as proved by the comparison with Hg-1223,
Cu-1212 or Cu-2212 systems. The detailed analysis of the atomic
positions shows that the shrinkage of the ``(Ba/Sr)O$_{9}$''
coordination polyhedra with the Sr content, i.e. the strong
decrease of the Ba/Sr-O(1)(2)(3) bonds. As a consequence, Cu-O1
and Ba/Sr-O2 planes become closer, but no effect on the Y-O1 bonds
(i.e. between the CuO$_{2}$ superconducting planes) is observed.
The Sr chemical pressure has a pronounced effect on the Hg/Y-O3
bond which increases, making the Hg/Y-O flat, while it decreases
strongly the Cu-O1$_{in-plane}$ and Cu-O2$_{apical}$ bonds. These
structural changes, and particulary the two last ones, should
affect greatly the charge repartition in the 2212 lattice and then
the charge transfer towards the superconducting planes, explaining
the observed $T_{c}$ increase with the Sr content. This aspect
will be developed in the part 2 of this article using a "bond
valence sum" approach.

\ack P. Toulemonde thanks CNRS for its financial support during
its PhD research. The authors acknowledge R. Argoud, C. Brachet
and R.~Bruy\`{e}re for technical assistance in HP-HT experiments.
The authors are grateful to stimulating discussions with Dr J. L.
Tholence and his help in measuring the a.c. susceptibilities.

\end{document}